\documentclass[aps, prd, floatfix, nofootinbib, superscriptaddress, twocolumn]{revtex4-1}

\usepackage[utf8]{inputenc}
\usepackage[brazil]{babel}
\usepackage[T1]{fontenc}
\usepackage{latexsym}
\usepackage{amsmath}
\usepackage{amssymb}
\usepackage{amsfonts}
\usepackage[mathscr,scaled=1.15]{urwchancal}
\DeclareFontFamily{OT1}{pzc}{}
\DeclareFontShape{OT1}{pzc}{m}{it}%
{<-> s * [1.15] pzcmi7t}{}
\DeclareMathAlphabet{\mathpzc}{OT1}{pzc}{m}{it}

\usepackage{color}

\usepackage{supertabular}
\usepackage{placeins}
\usepackage{epsfig}
\usepackage{graphicx}

\definecolor{purple}{rgb}{0.5,0,0.5}
\definecolor{blue}{rgb}{0.0,0,0.9}
\definecolor{prdblue}{rgb}{0.133,0.118,0.498}
\usepackage[colorlinks=true, pdfstartview=FitV, linkcolor=prdblue, citecolor= prdblue, urlcolor=prdblue]{hyperref}

\begin{document}

\title{O uso do Arduíno como uma ferramenta metodológica para o ensino de física no ensino médio \\ {\small The use of Arduino as a methodological tool for teaching physics in high school}}

\author{Ítalo M. de Lima}
\email{italoroci@hotmail.com}
\affiliation{\mbox{Universidade Federal do Vale do São Francisco}, Juazeiro, BA, Brasil}
\affiliation{\mbox{Instituto Federal de Educação, Ciência e Tecnologia do Piauí, Grupo de Pesquisa em Ensino de Física},\\ Picos, PI, Brasil}

\author{Emanuel V. de Souza} 
\email{emanuel.veras@ifpi.edu.br}
\affiliation{\mbox{Instituto Federal de Educação, Ciência e Tecnologia do Piauí, Grupo de Pesquisa em Ensino de Física},\\ Picos, PI, Brasil}

\author{Francisco D. V. de Araújo} 
\email{fdiasisva@gmail.com}
\affiliation{\mbox{Instituto Federal de Educação, Ciência e Tecnologia do Piauí}, Campo Maior, PI, Brasil}


\date{23 março 2020}
%\date{01 February 2019}

\begin{abstract}
Tendo em vista as dificuldades encontradas no ensino de ciências, especificamente no ensino de física, estudos visam desenvolver metodologias que auxiliem o ensino de ciências. Neste âmbito, o presente trabalho propõe a utilização da plataforma Arduíno como uma ferramenta metodológica para aperfeiçoar o ensino de física no ensino médio. Com esse propósito, o objetivo do trabalho foi produzir experimentos de física com o auxílio do Arduíno que podem ser utilizados em sala de aula e desse modo investigamos que os experimentos proporcionaram aos alunos uma aula dinâmica e próxima do ideal no ensino de física. Para o desenvolvimento da pesquisa foram apresentados dois experimentos, nas áreas de termologia e óptica, para alunos da segunda série de ensino médio de duas escolas públicas da cidade de Picos (PI). Após o desenvolvimento da atividade experimental, foi aplicado um simples questionário que visou embasar se os experimentos utilizados com o Arduíno contribuiu para o ensino de física. Os resultados obtidos foram discutidos a partir do questionário aplicado chegando ao indicativo de que os experimentos produzidos com o Arduíno contribuíram significativamente para o ensino aprendizagem da física, indicando que é uma ferramenta metodológica excelente para o ensino de física e ciências em geral.\\
\textbf{Palavras-chaves}: Tecnologia da informação e comunicação; Experimentos de física; Arduíno; Ensino-Aprendizagem.  \\

In view of the difficulties encountered in science teaching, specifically in the teaching of physics, studies aim to develop methodologies that help science teaching. In this context, the present work proposes the use of the Arduino platform as a methodological tool to improve the teaching of physics in high school. For this purpose, the objective of the work was to produce physics experiments with the aid of Arduino that can be used in the classroom and thus we investigated that the experiments provided students with a dynamic and close to ideal class in physics teaching. For the development of the research, two experiments were presented, in the areas of thermology and optics, for students of the second grade of high school from two public schools in the city of Picos (PI). After the development of the experimental activity, a simple questionnaire was applied that aimed to support whether the experiments used with Arduino contributed to the teaching of physics. The results obtained were discussed based on the questionnaire applied, indicating that the experiments produced with Arduino contributed significantly to teaching physics learning, indicating that it is an excellent methodological tool for teaching physics and science in general. 
\end{abstract}

\maketitle

%%%%%%%%%%%%%%%%%%%%%%%%%%%%%%%%%%%%%%%%%%%%%%%%%%%%%%%%%%%%%%%%%%%%%%%%%%%%%%%%%%%%%%%%%%%%%%%%%%%%%%%%%%%%%%%%%%%%%%%
% 4500 words

%\noindent\textbf{1.$\;$Introduction}.
\section{Introdução}

As ciências exatas e da natureza (Física, Química, Biologia, Matemática e Tecnologias), no mundo atual, se fazem necessárias para o desenvolvimento da sociedade tanto no quesito tecnológico quanto no social. No Brasil, infelizmente, o ensino dessas ciências se encontra defasado, sendo ensinada de maneira ainda desprovida de contextualização e significado, tornando para os alunos disciplinas complexas e as vezes muito difícil de se aprender, sem ao menos eles perceberem que essas disciplinas estão extremamente inseridas na nossa vida cotidiana~\cite{Praxedes}. Nesse contexto surge a necessidade de estudar e desenvolver novos métodos para melhorar o ensino aprendizagem destas ciências, em especial a física, fazendo com que os discentes vejam a disciplina de uma maneira que desperte neles a curiosidade de investigar, do querer entender e desenvolver o mundo em que vivem. Os autores Araújo e Abib~\cite{Araujo} relatam em seu trabalho que as dificuldades e problemas, que tanto afeta o ensino de física, vem sendo estudadas a bastante tempo e levando a diferentes reflexões sobre a solução desse problema. 

Um dos pioneiros na busca de novas técnicas e métodos de ensino de ciências foram os EUA (Estados Unidos da América), que pós-segunda guerra mundial iniciou a corrida espacial contra a Ex-URSS (União das Repúblicas Socialistas Soviéticas) quando a mesma lançou o satélite Sputnik na década de 1950. Para não ficarem atrás, os EUA começaram a pesquisar e desenvolver técnicas para deixar a ciência mais atraente e interessante e com isso formar novos cientistas e engenheiros. As técnicas que mais se destacaram nesse processo foram à argumentação e o uso de experimentos, sendo a segunda a mais utilizada~\cite{Carvalho}. Sobre a importância de se utilizar a experimentação, Araújo e Abib~\cite{Araujo} enfatiza que 

\begin{quote}
“[...] o uso de atividades experimentais como estratégia de ensino de Física tem sido apontado por professores e alunos como uma das maneiras mais frutíferas de se minimizar as dificuldades de se aprender e de se ensinar Física de modo significativo e consistente ”. (Pág. 176)
\end{quote}

A prática experimental nas aulas de física se faz necessária, pois a mesma se trata de uma disciplina teórica-experimental, logo a falta de experimentação tornará o ensino de física ineficiente e dificultoso. Dessa forma a física se torna, para os discentes, uma disciplina baseada apenas em fórmulas e gráficos se distanciando do seu objeto de estudo que é a discussão e observação dos fenômenos da Natureza e o desenvolvimento tecnológico. 

Ensinar não é apenas transmitir conteúdo, mas é criar possibilidades do aluno produzir seu próprio conhecimento e conceitos. Esse ensinamento é apresentado no livro Pedagogia do Oprimido do autor Paulo Freire~\cite{Freire}, um dos educadores mais notáveis na história da pedagogia mundial. Com base nesse ensinamento é de se esperar que os professores passem a utilizar a atividade experimental em sala de aula para que os discentes possam construir seu próprio conhecimento, se aproximando de fato do estudo de ciências, que é a investigação, a busca do entendimento e a construção do pesamento crítico e desse modo poder fortalecer a Ciência no Brasil.

No Brasil, pelo baixo investimento financeiro nas escolas públicas e particulares, em sua maioria, não possuem laboratórios de física, pois estes possuem um alto custo de aquisição, o que dificulta a solução do problema no ensino de física diagnosticado pela maioria dos pesquisadores como aponta Araújo e Abib~\cite{Araujo}. É nesse contexto que se insere o grande avanço tecnológico dos últimos anos principalmente na computação. Mokarzel e Soma~\cite{Mokarzel} relatam que o extraordinário desenvolvimento dos sistemas computacionais eletrônicos digitais ocorreu na segunda metade do século XX, e que tem se intensificado ainda mais nos dias atuais. E com esse avanço tecnológico foram desenvolvidas ferramentas que auxiliam a educação, tais ferramentas são chamadas de TIC (Tecnologias de Informação e Comunicação). Correia~\cite{Correia} descreve em seu trabalho o conceito e a importância das TIC na sociedade e na educação,

\begin{quote}
“As TIC constituem não só uma ferramenta ao serviço do processo de ensino-aprendizagem, mas principalmente um instrumento que propicia representar e comunicar o pensamento, atualizá-lo continuamente, resolver problemas e desenvolver projetos. A utilização das TIC favorece a articulação entre as diversas áreas do saber, proporcionando um aprofundamento de alguns conteúdos específicos e levando à produção de novos conhecimentos”. (Pág. 8)
\end{quote}

Diante de novas tecnologias no mundo atual é de se esperar que a educação se aproprie destas tecnologias e use a seu favor. Sendo assim, em vista da rapidez do crescimento e avanço destas tecnologias, professores de física devem acompanhar esse ritmo e usar cada vez mais as TIC como afirmam os autores da Ref.~\cite{Dourado} relatando que “as TIC tornam-se um meio de integração entre o professor e o aluno, buscando novas metodologias para inovar a maneira de ensinar e aprender, no sentido de promover a interação entre o aluno e o novo cenário onde estão inseridos, no contexto do mundo atual” (Pág. 359). Para complementar essa afirmação, segundo Júnior~\cite{Junior}, o uso de atividades experimentais aguçará a curiosidade do aluno levando o mesmo a investigar os fenômenos físicos e desse modo construir o conhecimento significativo. 

Tendo em vista o alto custo para a aquisição de equipamentos para a montagem de laboratórios de física, uma ótima alternativa para suprir a necessidade das atividades experimentais em escolas públicas e particulares, com limitações financeiras, são os equipamentos eletrônicos que se adaptam aos computadores, por meio de placas, para as mais diversas funções e que possuem o baixo custo de aquisição. Dentre essas placas, temos em destaque o Arduíno, uma plataforma de prototipagem eletrônica formada por hardware e software livres, de baixo custo, e que pode ser usada em sistemas de aquisição de dados, ou seja, na construção de experimentos de física de baixo custo. 

O Arduíno trata-se de uma placa que interage o ambiente virtual com o ambiente físico por meio de sensores ligados a portas digitais e analógicas. O mesmo possui uma linguagem de programação própria, baseada em C++, linguagem essa muito conhecida e utilizada dentre as linguagens de programação, o que torna o Arduíno uma ferramenta útil para professores, pois a mesma não exige o estudo de uma nova linguagem de programação e eletrônica e em contrapartida oferece uma gama de possibilidades na criação de experimentos de física. 

Alguns trabalhos já nos trazem bons indicativos do uso do Arduíno para a prática do ensino de física. Os autores da Ref.~\cite{Souza}, por exemplo, abordam o uso do Arduíno como uma opção de baixo custo para realizar experimentos de física que podem ser aplicados em sala de aula. No respectivo trabalho, a pesquisa foca o uso do Arduíno no estudo didático de oscilações amortecidas em circuitos elétricos. Podemos também encontrar outras aplicações do uso do Arduíno para prática de experiências de física, como o estudo do tempo de carga e descarga de um capacitor em um circuito RC~\cite{Cavalcante} e a medida da constante gravitacional em um experimento de queda livre~\cite{Cordova}. Todas as propostas apresentadas, nos trabalhos citados anteriormente, podem ser usados tanto na abordagem do ensino médio como no ensino superior. Assim, o Arduíno constitui uma ferramenta auxiliar para educação, em especial para o ensino de física, pois o mesmo se faz como um laboratório portátil para professores, abrangendo toda a física, desde a mecânica até a física moderna. Deste modo, o objetivo deste trabalho é propor uma ferramenta metodológica para o ensino de física, ou seja, utilizar o Arduíno para produzir experimentos de baixo custo, e dessa forma melhorar a compreensão dos alunos por meio desses experimentos aproximando os discentes da verdadeira ciência, bem como investigar se o mesmo funcionará de forma significativa para auxiliar o ensino-aprendizagem da física.

Nesse contexto, investigou-se o uso da placa Arduíno em práticas experimentais de física e analisou-se o quanto essa inserção no ensino de física é eficiente no lugar de laboratórios ausentes nas escolas públicas e particulares. Para uma melhor apresentação, este trabalho está organizado da forma como segue. Na seção 2, descrevemos de forma mais detalhada a placa Arduíno, apresentando os principais conceitos sobre seu hardware, software e componentes necessários para a utilização da placa e para produção de experimentos para o ensino de física. Na seção 3, mostramos algumas aplicações do Arduíno nas ciências, demostrando o leque de aplicações que o mesmo oferece e as mais diversas possibilidades de se utilizar desses experimentos em sala de aula. Já na seção 4, apresentamos a metodologia utilizada neste trabalho, mostrando os experimentos construídos e usados durante a pesquisa. Demonstra-se também, como foram aplicados esses experimentos, além de discutir brevemente os conteúdos abordados durante aplicação dos experimentos. E por fim, nas seções 5 e 6, apresentamos, respectivamente, a discussão dos resultados e a conclusão da pesquisa, bem como algumas perspectivas de futuros trabalhos.       

%%%%%%%%%%%%%%%%%%%%%%%%%%%%%%%%%%%%%%%%%%%%%%%%%%%%%%%%%%%%%%%%%%%%%%%%%%%%%%%%%%%%%%%

\section{Plataforma Arduíno}

\begin{figure}[!h]
\centering
\includegraphics[scale=0.14]{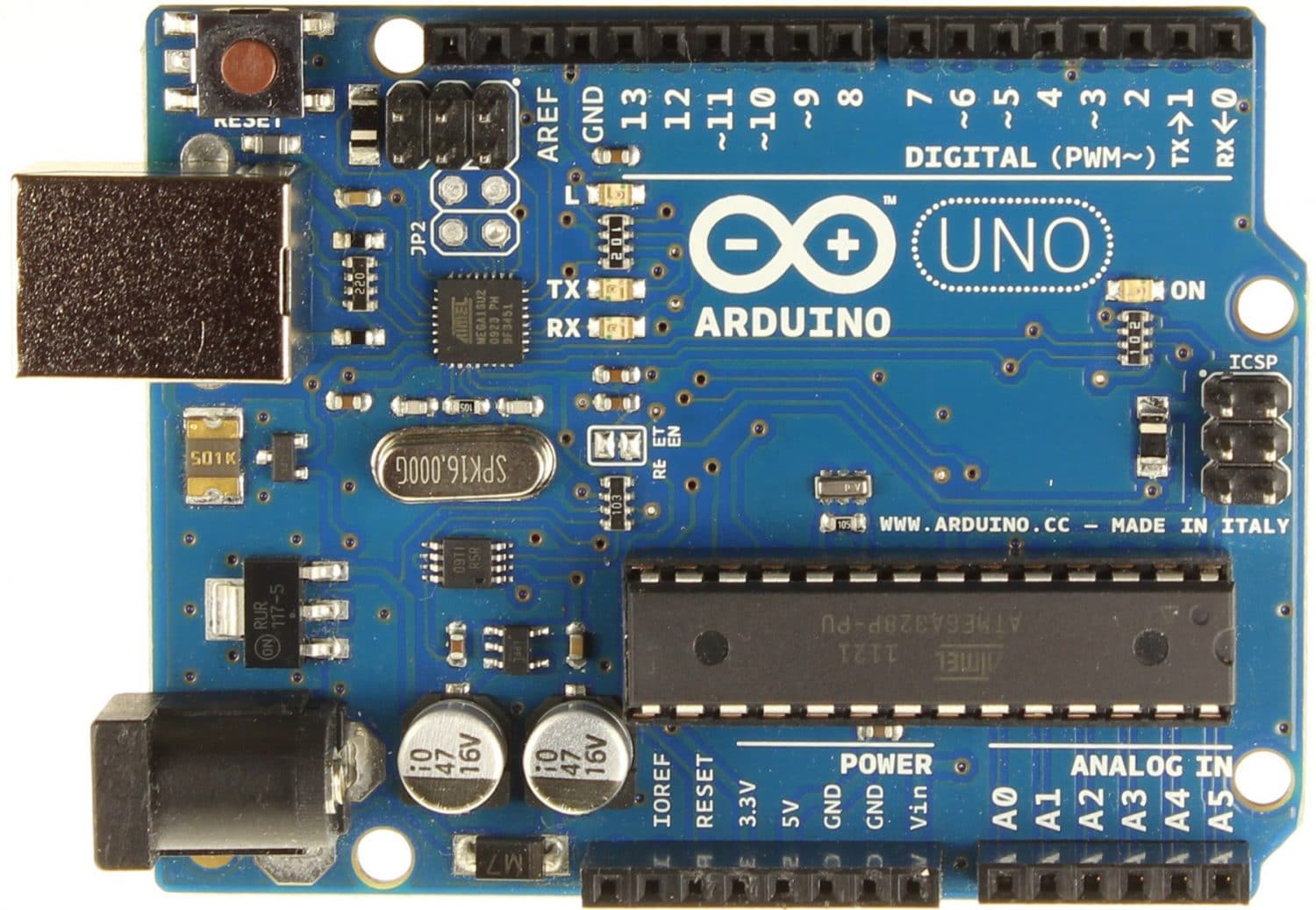}
\caption{Placa Arduíno UNO Rev. 3. A placa possui dimensões $\sim$ 5,3 cm x 6,8 cm x 1,0 cm. (Imagem disponível publicamente em \href{https://www.embarcados.com.br/placas-arduino/}{https://www.embarcados.com.br/placas-arduino})}
\label{arduino}
\end{figure}

A plataforma Arduíno teve sua origem na cidade de Ivrea, Itália, no ano de 2005, produzida por um grupo de cinco pesquisadores que tinham como objetivo construir um dispositivo que permitisse a criação de projetos eletrônicos de baixo custo, com o propósito de ser acessível a todos os tipos de usuários, incluindo estudantes, amadores, profissionais e/ou aficionados por robótica. Na Fig.~\ref{arduino} podemos observar os aspectos visuais do modelo mais atual da placa, o Arduíno UNO Rev. 3, após um aperfeiçoamento dos primeiros modelos desenvolvidos. Para o desenvolvimento da pesquisa, usamos o mesmo modelo apresentado na Fig.~\ref{arduino}.

%%%%%%%%%%%%%%%%%%%%%%%%%%%%%%%%%%%%%%%%%%%%%%%%%
%%%%%%%%%%%%%%%%%%%%%%%%%%%%%%%%%%%%%%%%%%%

Em termos técnicos, a plataforma Arduíno é uma placa de prototipagem com hardware e software livres~\footnote{Possibilidade de qualquer pessoa modificar e/ou melhorar o Arduíno a partir do hardware ou software básico.}, placa única e com linguagem de programação própria baseada em C++. Também possui um microcontrolador ATMEL~\footnote{Microcontrolador com memória flash.} e circuitos de entrada e saída de sinais embutidos. A mesma é programada a partir de uma IDE (Integrated Development Environment\footnote{ Em tradução livre significa "Ambiente de Desenvolvimento Integrado".}) de programação baseada em Processing~\footnote{Linguagem de programação de código aberto voltado para as artes eletrônicas e comunidades de projetos visuais.}, que consiste no ambiente de programação do Arduíno.

De acordo com Mcroberts~\cite{Mcroberts}, “o Arduíno é o que chamamos de plataforma de computação física ou embarcada, ou seja, um sistema que pode interagir com seu ambiente por meio de hardware e software” (Pág. 22).  De modo geral, a placa Arduíno é um microcomputador em que você pode programar, unindo hardware, software e o ambiente real, por meio de entradas analógicas e digitais ligadas a sensores. 

Um exemplo simplório do uso da placa Arduíno consiste na utilização da placa, de um sensor de temperatura, de jumpers (pequeno fio condutor) e uma fonte de energia (tais componentes serão descritos com mais detalhes nas próximas seções). Podemos programar o Arduíno para medir a temperatura de um ambiente e com isso utilizar esse protótipo em vários ambientes distintos, além de utilizá-lo com objetivos educacionais ou de automação. Como exemplo de automação, podemos citar um projeto que pode ser utilizado para controlar a temperatura de uma sala ou ambiente, em que o Arduíno é programado para captar a temperatura e limitar o aumento ou a diminuição da mesma, desligando ou ligando a central de ar quando necessário. Pode-se ainda, utilizando o mesmo princípio, usar o protótipo para monitoramento de uma granja ou ainda no solo agrícola, usando nesse caso não o sensor de temperatura, mas sim um sensor de umidade do solo. Com isso podemos monitorar o ressecamento do solo e fazer ligar ou não o sistema de irrigação. 

O fato da plataforma Arduíno ter hardware e software livre facilita a criação e o desenvolvimento de outras placas seguindo o mesmo padrão, porem apresentando algumas diferenças no seu hardware, software e alguns outros componentes. Nas subseções seguintes descrevemos com mais detalhes o Arduíno UNO Rev. 3 no que tange o seu hardware, software e componentes, conhecendo ainda melhor as funções e potencialidades da plataforma Arduíno. 

\subsection{Hardware do Arduíno UNO Rev. 3}

\begin{figure*}[!]
\centering
\includegraphics[scale=0.6]{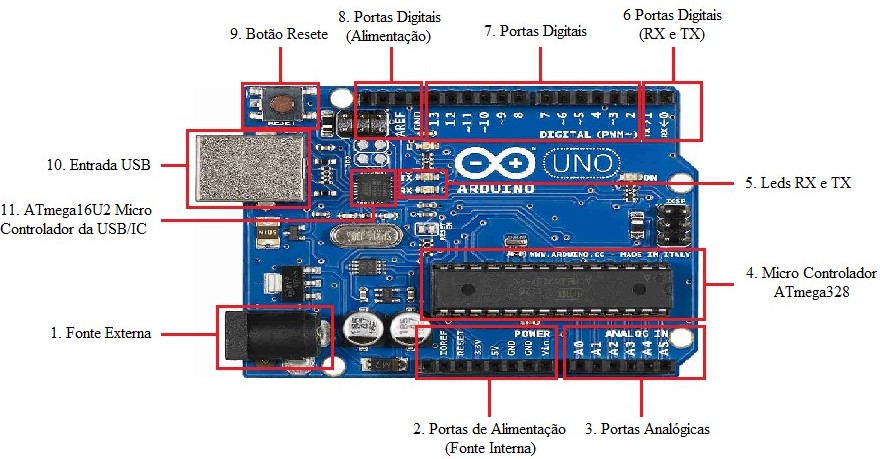}
\caption{Componentes da placa arduino UNO Rev. 3.}
\label{arduino2}
\end{figure*}

A placa Arduíno UNO Rev. 3 consiste no modelo mais popular e atual da plataforma. A mesma teve algumas atualizações desde sua criação que modificou entre outras características a capacidade de armazenamento, o processamento de dados e a quantidade de portas digitais e analógicas. Uma breve história da evolução da plataforma Arduíno pode ser encontrada na Ref.~\cite{arduino}. Entre as suas características principais o Arduíno UNO Rev. 3 possui catorze portas digitais, seis portas analógicas, um microcontrolador Atmega328, conexão USB entre outros componentes como podemos observar em detalhes na Figura~\ref{arduino2}.

%%%%%%%%%%%%%%%%%%%%%%%%%%%%%%%%%%%%%%%%%%%%%%%%%
%%%%%%%%%%%%%%%%%%%%%%%%%%%%%%%%%%%%%%%%%%%

A seguir descrevemos de forma mais detalhada os componentes do hardware do Arduíno UNO Rev. 3 mostrado na Figura~\ref{arduino2}. Em sua estrutura, podemos encontrar:

\begin{enumerate}
\item \textbf{Fonte de alimentação (externa)}: é uma porta que tem por finalidade receber a energia que irá alimentar o Arduíno, a mesma ainda tem como função receber tensão entre 7 e 12 V e fornecer na fonte de alimentação (interna) 3,3 V e 5 V.

\item \textbf{Portas de alimentação (interna)}: são as portas de saída de energia, ela fornece tensões de 3,3 V e 5 V, além do GND\footnote{Abreviação de Ground. Em circuitos eletrônicos faz referência a tensão nula (0 V). Normalmente está associado ao polo negativo das baterias.} (Terra). São nessas portas onde os sensores, shield\footnote{São placas de hardware que podem ser anexadas (plugadas) no Arduíno para aumentar a funcionalidade da placa.} e todo o circuito externo são alimentados energeticamente.

\item \textbf{Portas analógicas}: o Arduíno possui 6 portas analógicas que têm como função fazer a leitura de dados contínuos, ou seja, ela pode ter/ler qualquer valor no decorrer do tempo. Dessa forma essas portas são úteis na captação de dados, como temperatura, luminosidade, umidade, pressão e velocidade, pois essas grandezas não podem assumir apenas dois valores. Como exemplo, a temperatura não pode estar apenas no 0 ºC ou 100 ºC, no decorrer do dia ela pode variar em muito valores além desses, dependendo do ambiente.

\item \textbf{Microcontrolador Atmega328}: o Atmega328 é um computador completo, com memoria RAM e ROM além do núcleo de processamento dos dados. É nele onde são gravados e processados os programas e dados captados, o que dá ao Arduíno independência do computador e o torna tão versátil, além disso, o processador Atmega328 é quem interpreta os dados captados pelas diferentes portas e sensores, e conecta a máquina com o mundo real. 

\item \textbf{LEDs RX e TX}: mostram se as portas RX (receptor) e TX (transmissor) estão funcionando, ou seja, indicam entrada e a saída de sinais elétricos, respectivamente por meios das portas citadas, caso estejam funcionando como portas digitais, caso contrário, estão mostrando o funcionamento do programa com o envio e recebimento de dados entre o Arduíno e o computador.

\item \textbf{Portas RX e TX}: são portas de comunicação serial que podem funcionar como portas digitais, ou seja, receber ou emitir sinais, além de fazer a interação entre a placa Arduíno e o computador por meio da USB, ou com dispositivos portáteis, como a transmissão de dados via Bluetooth que permite controlarmos o Arduíno remotamente.

\item \textbf{Portas digitais}: no Arduíno UNO Rev. 3 temos 14 portas digitais de entrada e saída (as 6 portas analógicas podem funcionar como portas digitais, nesse caso temos 20 portas), que diferente das portas analógicas, conseguem ler apenas os valores lógicos bem definidos 0 volts e 5 volts. Dessa forma a função dessas portas é verificar, por exemplo se há um botão pressionado, ou se por meio de um sensor de presença se há ou não algo no ambiente onde o sensor esta inserido. De modo geral, essas portas representam a funcionalidade de um sensor, fazendo a leitura dos sinais captados pelo mesmo. 

\item \textbf{Portas de alimentação (interna)}: são as portas de saída de energia. Ela fornece tensões de 3,3 V e 5 V além do GND (Terra). São nessas portas onde os sensores, shield e todo o circuito externo são alimentos energeticamente.

\item \textbf{Botão reset}: na placa Arduíno UNO Rev. 3 não existe um botão que desliga a placa, para tal feito é necessário remover a fonte de energia interna. Com isso temos então o botão de reset que tem como finalidade reiniciar o programa. 

\item \textbf{Porta USB}: usada para fazer a comunicação do código binário (programa) entre o Arduíno e o computador. Também tem como função de alimentar energeticamente a placa Arduíno por meio do computador.

\item \textbf{Atmega16u2 controle do USB}: funciona como ponte da comunicação entre o Arduíno e o computador. É o responsável pela porta USB funcionar como serial, a mesma recebe o código e funciona como ponte entre o computador e o processador Atmega328.

\end{enumerate}

\subsection{Software do Arduíno UNO Rev. 3}

O IDE de programação (Software) do Arduíno, também conhecido como sketch, é o local responsável por criar os programas. O sketch possui um conjunto de comandos (instruções) lógicos que dirão ao arduino o que deve ser feito. Nesse caso as instruções definirão como serão captados e mostrados os dados obtidos pelos sensores. Logo após a criação desses programas é feito o upload dos mesmos na memória do Arduíno, deixando-o independente do computador \cite{Mcroberts}. 

\begin{figure}[!h]
\centering
\includegraphics[scale=0.6]{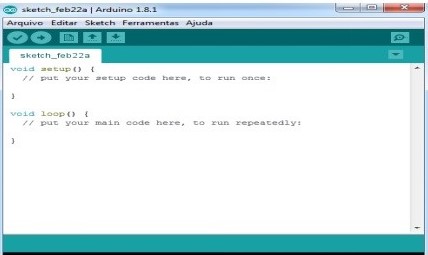}
\caption{O IDE de programação do Arduíno.}
\label{arduino3}
\end{figure}

A Figura~\ref{arduino3} mostra os aspectos visuais da plataforma de programação do Arduíno. Os códigos são programados em uma linguagem de programação baseada em C e C++. Com o intuito de facilitar o processo de programação para pessoas com pouco conhecimento nessa área, a interface de programação já vem com as duas principais funções da estrutura de programação do Arduíno, o \textit{Void Setup} e o \textit{Void Loop}.

%%%%%%%%%%%%%%%%%%%%%%%%%%%%%%%%%%%%%%%%%%%%%%%%%
%%%%%%%%%%%%%%%%%%%%%%%%%%%%%%%%%%%%%%%%%%%

O \textit{Void Setup} é uma função de entrada do programa. É nele onde definimos o nome do programa, variáveis, pinos de entrada de dados, entre outros. A execução desta função acontece apenas uma vez, para realizá-la novamente devemos usar o botão Reset, citado na subseção anterior, para reiniciar o programa.

O \textit{Void Loop} é a função onde está de fato o programa. Nela se define tudo o que o Arduíno deve captar por meio dos sensores, desde tempo de captação de cada dado até o momento de início e final do programa. É nessa função onde utilizamos as mais diversas estruturas de programação: estruturas de controle e os mais diversos operadores. Esta função é executada várias vezes até que o programa seja reiniciado ou até como tivermos programado. Vale ressaltar que não será discutido em detalhes sobre tais estruturas de controle e operadores do Arduíno, pois a ideia desta seção é de apresentar ao leitor os conceitos básicos sobre a plataforma Arduíno. Ao decorrer do artigo estaremos direcionando o leitor às diversas fontes bibliográficas para suprir essas necessidades.

O IDE de programação pode ser facilmente encontrada na Ref.~\cite{arduino2}, onde se tem a opção para diversos sistemas operacionais como: Linux, Mac OS e Windows. Ao fazer o download do programa, basta instalar e terá a plataforma como mostrado na Figura~\ref{arduino3}.

\begin{figure}[!h]
\centering
\includegraphics[scale=0.7]{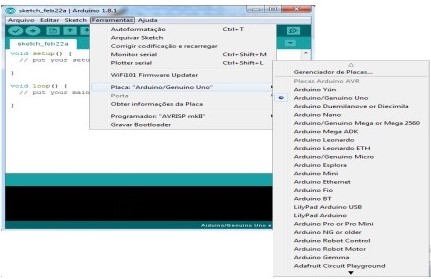}
\caption{Ferramenta no IDE de programação para escolha de qual Arduíno será utilizado no projeto.}
\label{arduino4}
\end{figure}

Para iniciar a programação, se faz necessário escolher qual Arduíno será utilizado, pois o IDE funciona para qualquer modelo de Arduíno, logo na interface de programação devemos indicar qual o Arduíno será utilizado, como podemos observar na Figura~\ref{arduino4}.

%%%%%%%%%%%%%%%%%%%%%%%%%%%%%%%%%%%%%%%%%%%%%%%%%
%%%%%%%%%%%%%%%%%%%%%%%%%%%%%%%%%%%%%%%%%%%

\begin{figure}[!h]
\centering
\includegraphics[scale=0.6]{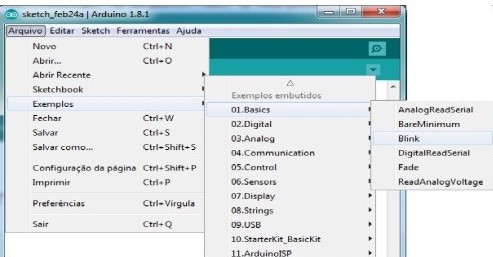}
\caption{Exemplo de projetos já pré-definidos no IDE de programação.}
\label{arduino5}
\end{figure}

Vale destacar que o exemplo mais simples e o primeiro a ser projetado com o arduino, semelhante ao “\textit{Hello World}” das linguagens de programação, consistem em acender um LED. O programa já está incluído na interface do Arduíno, junto com vários outros projetos. O “\textit{Hello World}” do Arduíno é chamado de Blink, como podemos observar na Figura~\ref{arduino5}. 

%%%%%%%%%%%%%%%%%%%%%%%%%%%%%%%%%%%%%%%%%%%%%%%%%
%%%%%%%%%%%%%%%%%%%%%%%%%%%%%%%%%%%%%%%%%%%

\subsection{Componentes para o arduino}

O Arduíno apresenta uma gama variada de componentes que são compatíveis com ele, que possibilitam a construção de diversos projetos, além de aumentar a capacidade do Arduíno o tornando uma placa tão versátil. Como exemplo podemos citar as shields, que são placas que ao serem conectadas no Arduíno ampliam a capacidade do mesmo, dando-lhe a função de se comunicar via Bluetooth ou conectar o Arduíno diretamente na rede de internet por WiFi, entre outras versatilidades. Alguns dos componentes usados no Arduíno são essenciais para determinados projetos. Nesta subseção iremos citar alguns dos mais utilizados, principalmente os usados na pesquisa.

\begin{figure}[!h]
\centering
\includegraphics[scale=0.36]{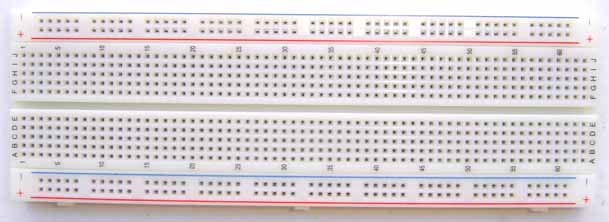}
\caption{Aspectos físicos da protoboard. Essa placa possui furos e conexões condutoras que ampliam a utilização do Arduíno e é comumente usada para experimentos com circuitos elétricos.}
\label{protoboard}
\end{figure}

A Figura~\ref{protoboard} mostra uma protoboard que é uma das peças componentes do projeto, a mesma tem como função de aumentar a capacidade do Arduíno de se conectar com mais componentes. Toda a protoboard pode ser alimentada a partir de dois pinos, 5 V e GND, gerando assim muitos espaços para conexões externas.

%%%%%%%%%%%%%%%%%%%%%%%%%%%%%%%%%%%%%%%%%%%%%%%%%
%%%%%%%%%%%%%%%%%%%%%%%%%%%%%%%%%%%%%%%%%%%

\begin{figure}[!h]
\centering
\includegraphics[scale=0.3]{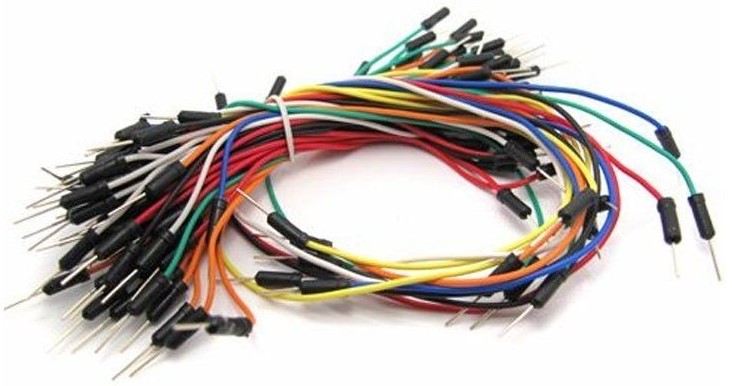}
\caption{Aspectos físicos de jumpers do tipo macho-macho.}
\label{jumpers}
\end{figure}

Na Figura~\ref{jumpers} mostramos os jumpers, que são os responsáveis pelas conexões de todo o projeto. É por meio deles que alimentamos a protoboard, sensores e demais ligações. Existem jumpers “fêmeas” e “machos”, o que permite conectar um “macho” a uma “fêmea” e deixar os sensores um pouco mais distante do Arduíno ou da protoboard.

%%%%%%%%%%%%%%%%%%%%%%%%%%%%%%%%%%%%%%%%%%%%%%%%%
%%%%%%%%%%%%%%%%%%%%%%%%%%%%%%%%%%%%%%%%%%%

\begin{figure}[!h]
\centering
\includegraphics[scale=0.6]{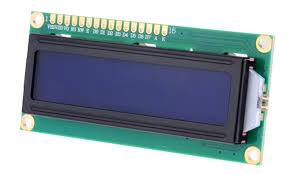}
\caption{Tela de cristal líquido do tipo 16x2 (2 linhas e 16 colunas).}
\label{lcd}
\end{figure}

Na Figura~\ref{lcd} mostramos outro componente bastante importante para o desenvolvimento dos experimentos, o LCD (Liquid Crystal Display). Esse componente nos dá um diferencial no projeto devido a independência que o mesmo fornece. Com ele podemos observar os dados coletados pelos sensores, como temperatura, umidade, luminosidade, data, hora entre outros deixando o projeto mais independente e completo.

%%%%%%%%%%%%%%%%%%%%%%%%%%%%%%%%%%%%%%%%%%%%%%%%%
%%%%%%%%%%%%%%%%%%%%%%%%%%%%%%%%%%%%%%%%%%%

\begin{figure}[!h]
\centering
\includegraphics[scale=0.35]{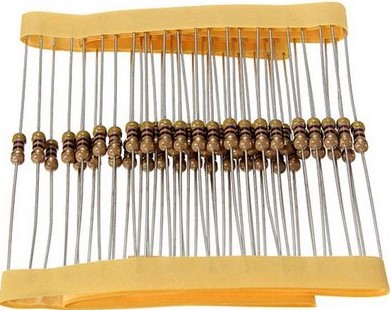}
\caption{Resistores de carbono de 1/4 W.}
\label{resistores}
\end{figure}

Já na Figura~\ref{resistores} mostramos os resistores para o uso com o Arduíno, pois tem como única função limitar a corrente que alimentará os sensores e/ou demais dispositivos, já que alguns dos dispositivos vêm com limitações de intensidade de corrente elétrica.

%%%%%%%%%%%%%%%%%%%%%%%%%%%%%%%%%%%%%%%%%%%%%%%%%
%%%%%%%%%%%%%%%%%%%%%%%%%%%%%%%%%%%%%%%%%%%

\begin{figure}[!h]
\centering
\includegraphics[scale=0.6]{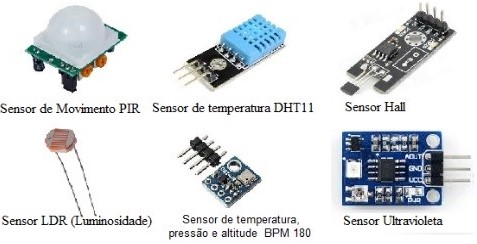}
\caption{Alguns tipos de sensores que podem ser utilizados em projeto de experimentos em ensino de física.}
\label{sensores}
\end{figure}

Na Figura~\ref{sensores} mostramos alguns sensores que são os responsáveis por captar os dados requeridos. Dentre eles, podemos observar o BPM 180 que foi utilizado neste trabalho e outros sensores que podem ser utilizados para produção de experimentos para o ensino de física.

%%%%%%%%%%%%%%%%%%%%%%%%%%%%%%%%%%%%%%%%%%%%%%%%%
%%%%%%%%%%%%%%%%%%%%%%%%%%%%%%%%%%%%%%%%%%%

\section{Uso da plataforma Arduíno no ensino de ciências}

O uso da plataforma Arduíno tem sido bastante comum desde sua criação em 2005, pois se trata de uma placa de prototipagem eletrônica de fácil utilização, com baixo custo de aquisição, além de ser versátil e ter grande aplicação. O mesmo é usado em diversas áreas do conhecimento, como por exemplo, no ensino e na pesquisa, através do desenvolvimento de protótipos de automação residencial, de segurança, industrial, entre outros, bem como em métodos que facilitem o ensino-aprendizagem da ciência. 

Os autores das Refs.~\cite{Filho, Lemos, Abe, Cury, Martinazzo} relatam em seus trabalhos sobre o uso do Arduíno no ensino de ciências e deixam claro a importância do mesmo. Iremos destacar a seguir experiências da utilização do Arduíno no ensino de física, biologia, química, informática/robótica e matemática, mostrando o leque de atuação da placa Arduíno como facilitador na aprendizagem do ensino de ciências. Vale destacar que as aplicações citadas nas subseções seguintes não são as únicas nem tão pouco somente nestas áreas.

\subsection{Arduíno no ensino de física}

Uma aplicação simples, porém, muito eficiente, consiste na utilização do Arduíno junto com um sensor de luminosidade LDR\footnote{O LDR (Light Dependent Resistor) consiste em um resistor que varia a sua resistência de acordo a intensidade luminosa.}, uma régua, um espelho e uma lanterna. Com esses materiais, temos a possibilidade de montar um experimento didático de física que está relacionado com as oscilações amortecidas, onde o sensor irá capturar a variação da intensidade luminosa da lanterna que estará oscilando como um pêndulo. A montagem do projeto é simples e de baixo custo, além de proporcionar um experimento que muito contribuirá para o entendimento dos conceitos de movimentos oscilatórios~\cite{Souza}. 

\begin{figure}[!h]
\centering
\includegraphics[scale=0.6]{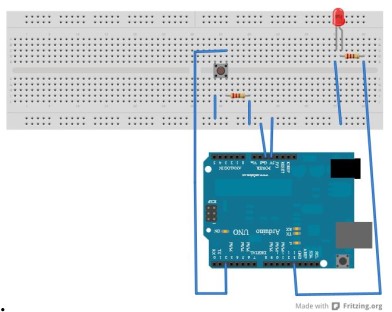}
\caption{Protótipo de um experimento para acender um LED. (Imagem retirada da Ref.~\cite{Fernandes})}
\label{prototipoled}
\end{figure}

Uma outra aplicação do Arduíno, desta vez, no ensino do conceito de eletricidade, tendo como objetivo de introduzir os principais conceitos sobre circuitos elétricos foi proposto pelos autores da Ref.~\cite{Fernandes}. No respectivo trabalho, os mesmos propõem a construção de enfeites natalinos com auxílio do Arduíno, mostrando aos alunos os conceitos sobre resistores e associação de resistores, capacitores, DDP, intensidade de corrente elétrica, entre outros conceitos. Um esquema da montagem do experimento pode ser observado na Figura~\ref{prototipoled}.

%%%%%%%%%%%%%%%%%%%%%%%%%%%%%%%%%%%%%%%%%%%%%%%%%
%%%%%%%%%%%%%%%%%%%%%%%%%%%%%%%%%%%%%%%%%%%

O Arduíno também pode ser utilizado para a construção de um experimento que faça a medida da aceleração da gravidade~\cite{Cordova}. Para a produção deste projeto, além do Arduíno, necessita-se de materiais bem simples de serem encontrados, como por exemplo, quatro barras de ferro e uma esfera de metal pequena. A esfera é presa entre duas barras de ferro a uma altura $H$ de uma base, que possui as outras duas barras de ferro, e ligadas ao Arduíno que capta uma tensão de cinco volts. No momento que se libera a esfera, a tensão vai a zero (na parte superior) e cai em uma base que funciona da mesma forma que as barras de ferro na parte superior do experimento, e a tensão que antes da esfera cair estava zero (na parte inferior) se torna cinco Volts quando a esfera toca essa base, sendo assim calculado o tempo de queda ($t$) e a aceleração da gravidade ($g$) por meio da relação $g=2H/t^2$, possibilitando o estudo do movimento de queda livre e grande parte da cinemática. Esta aplicação tem como foco principal os alunos do ensino médio, porém pode ser usado também na introdução de cursos superiores de física.

\subsection{Arduíno no ensino de biologia}

Na biologia, assim como nas demais ciências, há diversas possibilidades de se utilizar as TIC, de modo especial o Arduíno, que inserido no cenário atual, torna o processo educativo mais dinâmico, possibilitando maior aprendizagem e contribuindo para o tripé pesquisa, ensino e aprendizagem~\cite{Lemos}. 

	Um bom exemplo para utilizar o Arduíno no ensino de biologia é a construção de um protótipo para o estudo da fotossíntese em meio à variação da luminosidade em plantas aquáticas. Para essa produção, devemos utilizar diversos sensores como o de temperatura, qualidade do ar, luminosidade, pressão barométrica, jumpers, ethernet shield e uma protoboard. Com esse projeto há a possibilidade de captação de dados para análise e comparação com os dados de outros laboratórios, análise essa, poderá ser feita pelos os alunos, o que trará para os mesmos uma melhor noção sobre a pesquisa e análise de dados e um aprendizado mais significativo, além de proporcionar a interdisciplinaridade entre as ciências~\cite{Lemos}. 
	
\begin{figure}[!h]
\centering
\includegraphics[scale=0.65]{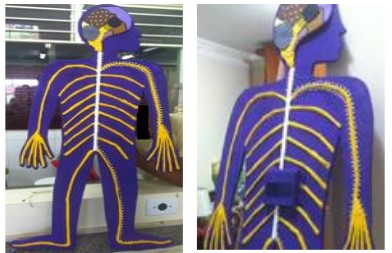}
\caption{Protótipo de um robô com o uso do Arduíno. (Imagem retirada da Ref.~\cite{Garcia})}
\label{maquete}
\end{figure}

	 Outra aplicação do Arduíno se faz no estudo de robótica educacional que possibilita a construção de conhecimentos de biologia por meio de robores, tendo como objetivo o estudo e desenvolvimento de conceitos da Biologia. Os autores da Ref.~\cite{Garcia} destacam a produção de um robô para o estudo do sistema nervoso central e periférico. Para a construção do mesmo, os autores relatam a utilização de materiais reutilizáveis como papelão, jornais e sucatas, além do Arduíno e seus componentes. Segundo os autores, a produção do robô foi feito por meio de um curso, com duração de 60 h, com alunos do ensino médio, onde inicialmente foram apresentados os principais conceitos sobre o Arduíno, seus componentes e sua programação. A Figura~\ref{maquete} mostra a maquete do experimento.
	 
%%%%%%%%%%%%%%%%%%%%%%%%%%%%%%%%%%%%%%%%%%%%%%%%%
%%%%%%%%%%%%%%%%%%%%%%%%%%%%%%%%%%%%%%%%%%%
	 
Com base na Ref.~\cite{Garcia}, a construção do robô foi realizada pelos alunos divididos em grupos, proporcionando assim o cooperativismo e a colaboração entre os mesmos. Para realizar o projeto, é necessário utilizar o Arduíno juntamente com LEDs, jumpers, dois botões e a maquete do robô (representação do corpo humano e seu sistema nervoso central). Os botões seriam os pés e as mãos e os LEDs o caminho percorrido pelo pulso nervoso.

Os autores da Ref.~\cite{Garcia} relatam a importância da utilização do Arduíno no ensino de biologia, segundo os mesmos, o uso dessa ferramenta bem como todas as TIC, faz com que o aluno tenha mais curiosidade, além de instigar os mesmos a fazerem reflexões e questionamentos sobre os conteúdos estudados.

\subsection{Arduíno no ensino de química}

\begin{figure}[!h]
\centering
\includegraphics[scale=0.5]{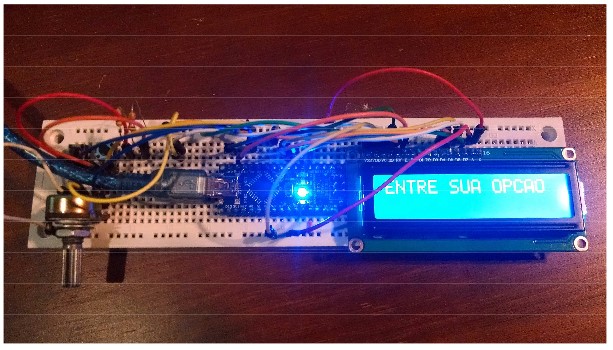}
\caption{Experimento para identificação de modelos dinâmicos de processos químicos. (Imagem retirada da Ref.~\cite{Filho})}
\label{esquemaqui}
\end{figure}

Nesta subseção mostramos como a placa Arduíno pode ser utilizada no ensino de química. A mesma pode ser usada, por exemplo, para a identificação de modelos dinâmicos de processos químicos, que visa observar o comportamento e o conhecimento desses processos baseados no ganho de processos no tempo. Para a produção do experimento é necessário a utilização do Arduíno, LCD, jumpers, protoboard, potenciômetros entre outros~\cite{Filho}. Podemos observar na Figura~\ref{esquemaqui} o esquema do experimento proposto.

%%%%%%%%%%%%%%%%%%%%%%%%%%%%%%%%%%%%%%%%%%%%%%%%%
%%%%%%%%%%%%%%%%%%%%%%%%%%%%%%%%%%%%%%%%%%%

Segundo o autor da Ref.~\cite{Filho}, o Arduíno foi usado na disciplina de controle de processos químicos do curso de Engenharia Química, e o uso do mesmo despertou grande curiosidade dos alunos e um maior interesse na disciplina, fazendo assim, com que a aprendizagem fosse mais significativa. O uso da plataforma dentro da disciplina, mostra que a desenvoltura e a facilidade do manuseio do Arduíno, que em pouco tempo (considerando uma disciplina de 60h) pode ser trabalhado de forma simples, porém muito significativa, faz com que professor e aluno adquiram um grande conhecimento no que tange o ensino aprendizagem. 

Outro exemplo da utilização do Arduíno no uso didático de química é a construção de um experimento para o estudo de espectroscopia. Segundo Moreira~\cite{Moreira}, o experimento denominado de fotômetro “demonstrou ser um instrumento confiável no registro de sinais analíticos com o propósito de determinar as concentrações de amostras de sulfato de níquel e, portanto, útil no ensino de espectrofotometria para alunos de Química”, tendo como objetivo comprovar a lei de Beer-Lambert. O experimento trabalha além da química, a física de maneira interdisciplinar, fazendo assim uma ótima ferramenta, não só para trabalhar o ensino por meio de experimentos, mas para elucidar a união dos conhecimentos de diversas ciências.

\subsection{Arduíno no ensino de informática/robótica}

O Arduíno também é destaque quando usado no ensino de informática. A placa facilita a interpretação da programação que é algo totalmente virtual e passa a ser algo virtual e prático com o auxílio do Arduíno. Isso é reafirmado por Abe~\cite{Abe} diz que, 

\begin{quote}
“A principal dificuldade no aprendizado da programação é o fato de programas de computadores serem objetos virtuais, exigindo que o estudante dessa área detenha uma alta capacidade de abstração para entender o que está sendo proposto. Assim, a utilização de um kit de componentes eletrônicos, juntamente com micro controladores mostra-se de grande ajuda na compreensão de conceitos básicos de programação, pois permite ao aluno observar fisicamente aquilo que foi programado, gerando um resultado palpável”. (Pág. 2)
\end{quote}

O Arduíno utiliza uma linguagem de programação baseada em C e C++, que são linguagens de destaque no mercado de trabalho, o que possibilita ao estudante uma boa aproximação dos principais conceitos desta linguagem amplamente utilizada~\cite{Abe}.

No ensino, o professor tem a oportunidade de utilizar o Arduíno de diversos meios. O trabalho dos autores da Ref.~\cite{Macedo} nos mostra diversas formas de aplicação do Arduíno como facilitador do ensino aprendizagem. No trabalho, os autores chamam a atenção para o uso do Arduíno em um curso de extensão, cujo objetivo é o ensino de disciplinas complexas como programação e arquitetura de computadores. Em outra aplicação~\cite{Macedo}, os autores relatam a experiência do curso de extensão intitulado “Aprendendo a Programar com LEDs”, onde se trabalhou a produção de um semáforo com o auxílio do Arduíno. Os autores destacam o grande aproveitamento do Arduíno como auxílio no ensino de informática e afirma que “os participantes estavam motivados por visualizarem o que foi programado por eles através do ambiente e dos dispositivos, compreendendo os conteúdos abstratos dessas disciplinas por meio de projetos práticos”(Pág. 4).

\begin{figure}[!h]
\centering
\includegraphics[scale=0.7]{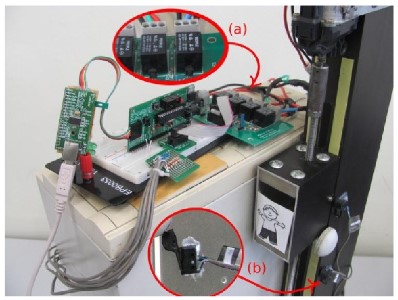}
\caption{Esquema de um elevador para prática didática de programação com o uso do Arduíno. (Imagem retirada da Ref.~\cite{Melo})}
\label{elevador}
\end{figure}

Outra forma de aplicarmos o Arduíno é descrito na Ref.~\cite{Melo}. Os autores utilizaram o Arduíno dentro da disciplina de programação I, no Curso Superior de Tecnologia em Mecatrônica Industrial, e relatam que obtiveram ótimos resultados com a utilização da metodologia. Na Figura~\ref{elevador} podemos observar o esquema de uns dos projetos desenvolvidos pelos autores.

%%%%%%%%%%%%%%%%%%%%%%%%%%%%%%%%%%%%%%%%%%%%%%%%%
%%%%%%%%%%%%%%%%%%%%%%%%%%%%%%%%%%%%%%%%%%%

\subsection{Arduino no ensino de matemática}

Sendo uma disciplina muito temida para alguns alunos, a matemática precisa de um dinamismo em suas aulas para facilitar o ensino desta disciplina tão difícil aparentemente. Cury e Hirschmann~\cite{Cury} enfatizam em seu trabalho a importância de ensinar matemática usando as tecnologias, ou seja, usando o que os alunos mais têm em sua realidade, fazendo uma aproximação entre o estudo abstrato da matemática com a realidade. Segundo os autores, o Arduíno pode ser utilizado no ensino de matemática, como por exemplo, para demostrar alguns conteúdos, como matrizes utilizando LEDs.

Outra forma de aplicação é usar o Arduíno para construir programas simples para obter soluções de problemas matemáticos que vão desde simples operações, como soma e subtração, a expressões algébricas, além disso, para o próprio estudo de linguagem de programação são necessárias ideias matemáticas, bem como raciocínio lógico~\cite{Cury}. 

Os resultados encontrados na Ref.~\cite{Cury} nos mostram que o uso do Arduíno no ensino de matemática irá despertar no aluno a criatividade, a autoestima e o raciocínio, se fazendo então uma ótima ferramenta metodológica para o ensino de matemática.

\section{Metodologia e aplicação em sala de aula}

Após apresentarmos em seções anteriores uma breve descrição sobre a placa Arduíno e a sua aplicabilidade no ensino, nesta seção iremos abordar sobre a prática metodológica do uso do Arduíno para a realização de experimentos em sala de aula.

O presente trabalho foi desenvolvido através da aplicação dos experimentos, nas áreas de termologia e óptica, em sala de aula, seguido da aplicação de um simples questionário e finalizado com a discussão teórica dos conteúdos ministrados. O questionário foi elaborado com questões objetivas e subjetivas que visa mensurar o quanto o experimento foi eficiente para a prática de ensino. As perguntas do questionário estavam fundamentadas nas seguintes questões,

\begin{itemize}

\item Se o aluno já tinha visitado algum laboratório de física e/ou realizado alguma prática experimental. Em caso afirmativo, onde o aluno realizou a visita e qual tipo de experimento o aluno já tinha realizado;

\item Se os experimentos realizados em sala de aula contribuíram de alguma forma para o entendimento dos conteúdo ministrados, especificando o quanto a prática contribuiu para isso;

\item Se atividade experimental produzida com Arduíno contribuiu de forma significativa para o entendimento e desenvolvimento da aula, explicando de que forma contribuiu para isso;

\item Com resposta subjetiva, o aluno deveria responder se o professor deveria utilizar a pratica experimental em suas aulas, e em caso afirmativo, especificando qual frequência;

\item Em relação ao Arduíno e com base nos conceitos apresentados e na experiência praticada, como o discente avalia a funcionalidade do mesmo para a produção de experimentos de física.

\end{itemize}

Antes da apresentação do experimento, foram mostrados os conceitos iniciais sobre o Arduíno, relatando as principais funcionalidades sobre a montagem, aplicações e sobre a linguagem de programação. Dessa forma, os alunos ficariam mais familiarizados com a prática. Em seguida foi fornecido aos discentes um roteiro dos experimentos (cf. os apêndices A e B) que iria guiá-los durante a prática. 

Os experimentos de termodinâmica e ótica, que serão discutidos nas próximas subseções, foram apresentados antes da abordagem teórica dos conteúdos correlacionados, assim esses conteúdos seriam estudados a partir dos experimentos. O intuito da apresentação nessa ordem, experimento e depois o conteúdo, tem como objetivo que o aluno faça uma análise sobre o experimento e crie hipóteses acerca das relações entre as grandezas físicas e os acontecimentos vistos, desenvolvendo assim um olhar crítico e científico sobre os fenômenos observados. Para a apresentação do experimento de termologia, além do arduino e seus componentes, foi utilizado uma garrafa com água em estado sólido, uma lâmpada ou a luz dos celulares e o próprio corpo humano, que seria usado para variar as grandezas captadas pelo sensor (temperatura, pressão e altitude). Para a apresentação do experimento de óptica foi utilizado além do Arduíno e seus componentes, uma lâmpada e duas semiesferas metálicas. 

Para o desenvolvimento da prática, os alunos foram divididos em 4 grupos, onde cada grupo teria que desenvolver hipóteses sobre o fenômeno analisado, ou seja, mostrassem de forma lógica as relações físicas e matemáticas entre as grandezas físicas observadas nos experimentos. Só após a apresentação do relatório com todas as informações, foi apresentado o conteúdo e corrigido possíveis dificuldades e equívocos sobre a teoria.

\subsection{O experimento de termologia}

\begin{figure}[!h]
\centering
\includegraphics[scale=0.07]{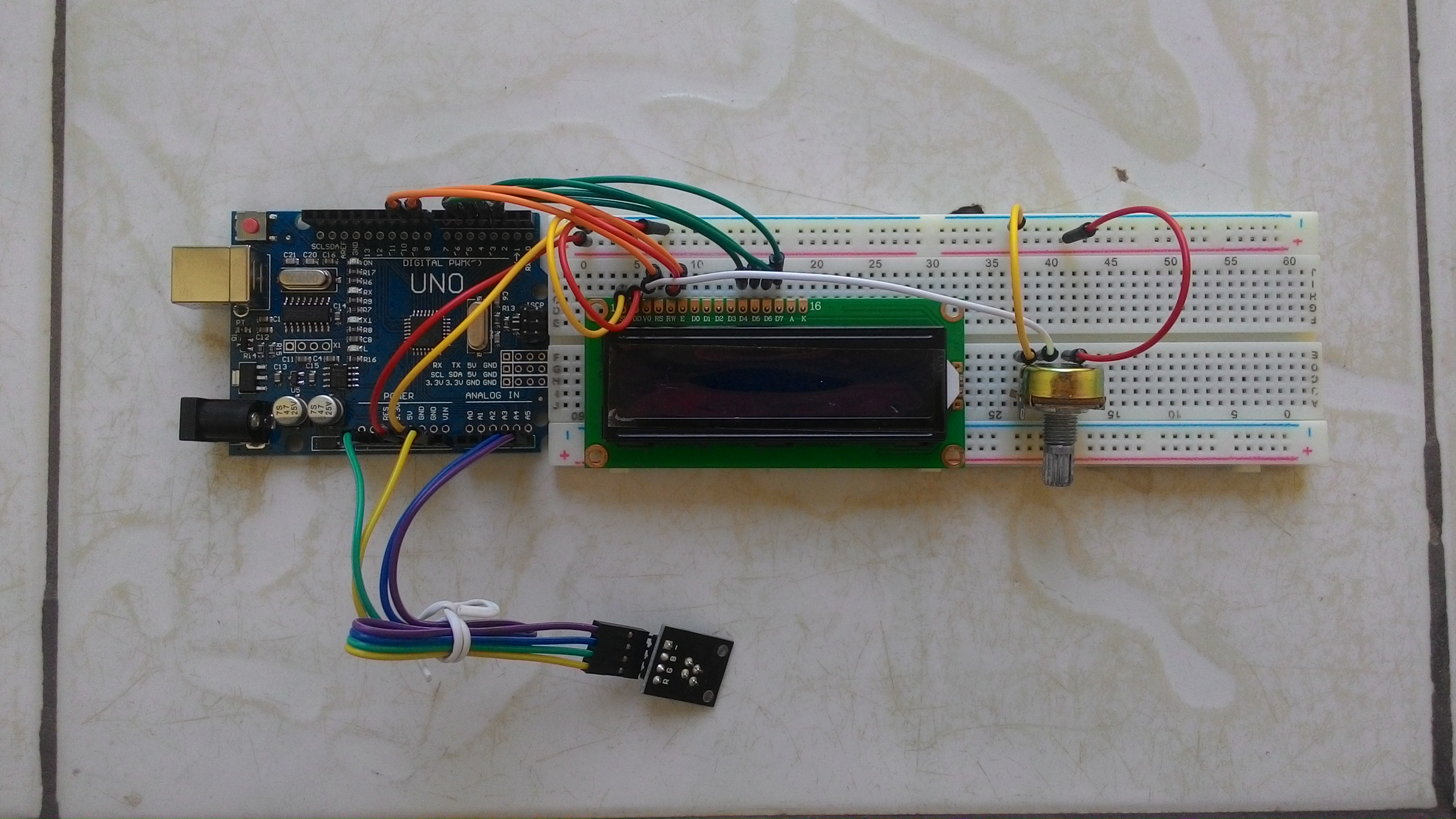}
\caption{Protótipo do experimento de termologia com o uso do arduino.}
\label{termologia}
\end{figure}

O experimento desenvolvido, denominado de termômetro, é apresentado na Figura~\ref{termologia}. O projeto foi construído com o objetivo de captar a temperatura, pressão atmosférica e altitude. Tais grandezas físicas são úteis para os estudos da termologia e termodinâmica, mostrando conceitos iniciais como temperatura, energia térmica, calor e as transformações gasosas. Ao observar a Figura~\ref{termologia}, o leitor deve estar se perguntando se de fato este simples protótipo fará a diferença no ensino aprendizagem, porém o trabalho tem como objetivo de apresentar o Arduíno na sua forma mais simples, e mostrar que algo simples como o experimento mostrado pode ou não chamar a atenção dos alunos.

%%%%%%%%%%%%%%%%%%%%%%%%%%%%%%%%%%%%%%%%%%%%%%%%%
%%%%%%%%%%%%%%%%%%%%%%%%%%%%%%%%%%%%%%%%%%%

A montagem do experimento é simples e consiste em utilizar o Arduíno UNO rev. 3 com auxílio de um sensor de temperatura, umidade e pressão BPM 180, dois potenciômetros, uma tela de LCD, jumpers, uma protoboard, uma fonte ou cabo USB. O esquema de montagem do experimento é mostrado na Figura~\ref{termologia}.

\subsection{O experimento de óptica}

\begin{figure}[!h]
\centering
\includegraphics[scale=0.07]{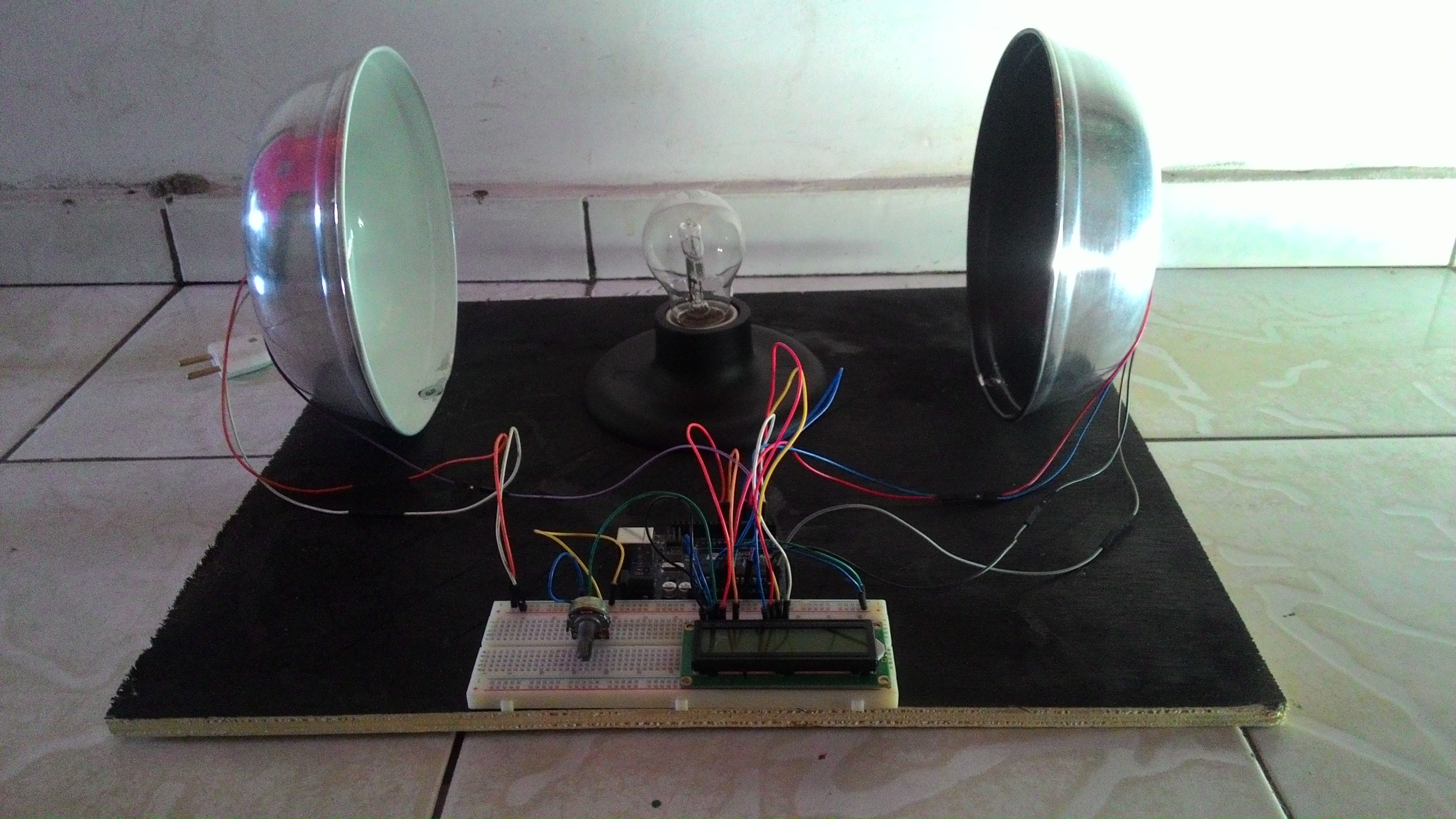}
\caption{Protótipo do Experimento de absorção da luz com o uso do Arduíno.}
\label{absorcao}
\end{figure}

O experimento de óptica, denominado de absorção luminosa, é mostrado na Figura~\ref{absorcao}. O mesmo tem como objetivo discutir os conceitos sobre absorção da luz em diferentes cores e de modo mais profundo, discutir a absorção radioativa e a sua relação entre o aumento de temperatura dos corpos a partir da transformação de energia radioativa em energia térmica. Com esse experimento é possível analisar a diferença entre o aumento de temperatura dos corpos de diferentes cores submetido à mesma fonte radioativa em função do tempo de emissão de radiação~\cite{Santos}.

%%%%%%%%%%%%%%%%%%%%%%%%%%%%%%%%%%%%%%%%%%%%%%%%%
%%%%%%%%%%%%%%%%%%%%%%%%%%%%%%%%%%%%%%%%%%%

A montagem do experimento é simples e consiste em utilizar o Arduíno UNO Rev. 3, com auxílio de dois sensores de temperatura, umidade e pressão BPM 180, dois potenciômetros, uma tela de LCD, jumpers, uma protoboard, uma fonte de energia ou cabo USB e duas semiesferas metálicas com diferentes cores na parte interna. Os sensores são acoplados na parte externa dessas esferas. Vale ressaltar que nesse experimento o sensor BPM 180 será utilizado apenas com a função de captação de temperatura. O protótipo do experimento é mostrado na Figura~\ref{absorcao}.

Ressaltamos que não será apresentado o código fonte, ou seja, o programa do Arduíno dos experimentos apresentados, pois o objetivo da pesquisa não é fornecer um tutorial, mas mostrar que a utilização do experimento com o auxílio do Arduíno pode contribuir de forma significativa para as aulas e aprendizagem dos alunos. Caso o leitor queira mais detalhes sobre o código dos experimentos apresentados, pode entrar em contato com os autores.

\section{Resultados e discussão}

A pesquisa foi realizada com alunos do segundo ano do ensino médio de duas escolas da rede estadual de ensino da cidade de Picos (PI) com o intuito de obter respostas quanto ao questionamento sobre a importância da utilização de experimentos de física no ensino, dando ênfase nos experimentos preparados com auxílio do Arduíno. Obtivemos através da pesquisa um montante de 52 questionários, no qual os entrevistados possuíam uma faixa etária entre 15 a 18 anos.

Para avaliar a necessidade de se utilizar experimentos de física durantes as aulas, inicialmente foi questionado se os alunos já haviam conhecido algum laboratório de física ou se haviam feito alguma prática experimental em laboratório. Os dados coletados mostraram que mais da metade dos alunos, 53\%, não tiveram a oportunidade de visitar ou conhecer um laboratório de física. Vale destacar que os 47\% dos entrevistados que relataram que foram a algum laboratório, dizem ter ido ao laboratório de física do IFPI/\textit{Campus} Picos, durante um evento promovido pela própria instituição ou em uma única feira de ciências realizada por alunos do PIBID (Programa Institucional de Bolsas de Iniciação à Docência) também do IFPI/\textit{Campus} Picos, ou seja, nas escolas onde estudaram e estudam não tiveram a oportunidade de desenvolver atividades experimentais em laboratórios. Esses dados são indicativos da pouca utilização da atividade experimental nas escolas pesquisadas e da falta de estrutura de laboratórios de física das mesmas, o que torna o Arduíno uma excelente ferramenta para professores e/ou escolas, já que se trata de um instrumento de baixo custo e com uma gama muito grande de possibilidades para construção de atividades experimentais, além de fácil utilização. 	   

Em seguida perguntou-se aos alunos se os mesmos já haviam ou não realizado alguma atividade experimental, fora e/ou dentro do ambiente escolar. Os dados coletados mostraram que 60\% dos alunos já participaram que alguma atividade, enquanto 40\% não realizaram nenhuma prática experimental. Neste caso, a pesquisa mostra que boa parte dos alunos realizaram alguma atividade experimental.

Ao observar os resultados apresentados, percebemos que houve um aumento em relação aos alunos que conhecem algum laboratório de física e os que realizaram alguma atividade experimental de física. Esse aumento nos dá uma validade maior dos dados seguintes, tendo em vista que os alunos já tiveram algum contato com atividades experimentais e dessa forma tem poder de avaliar ainda melhor a necessidade dos mesmos, bem como o quanto auxiliam na aprendizagem.

\begin{figure}[!h]
\centering
\includegraphics[scale=0.63]{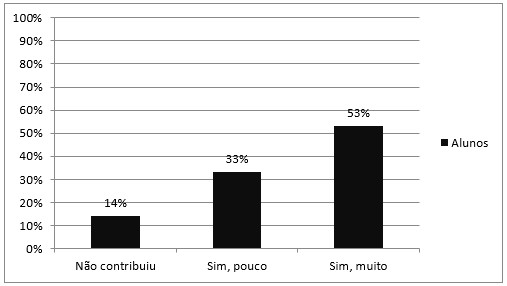}
\caption{Percentual de respostas quanto a contribuição do experimento para a aprendizagem dos temas abordados em sala de aula.}
\label{grafico1}
\end{figure}

Foi realizado então o terceiro questionamento, que diz respeito ao quanto o experimento realizado contribuiu para a aprendizagem dos alunos. Através do gráfico da Figura~\ref{grafico1}, observamos que 86\% dos entrevistados, ou seja, a maioria dos alunos, disseram que o experimento realizado contribuiu de forma significativa para aprendizagem dos conteúdos apresentados. O gráfico apresentado mostra três opções, sobre o quanto o experimento contribui para compreensão dos temas propostos, dentre os que responderam que sim, o experimento contribui, temos uma grande diferença daqueles que afirmam que o Arduíno contribuiu muito para aprendizagem dos mesmos e aqueles que dizem ter contribuído, porem pouco, onde percebemos que mesmo entre estes o experimento se faz necessário segundo os entrevistados. Estes resultados estão de acordo com os resultados da Ref.~\cite{Araujo}, que relata a importância da utilização das atividades experimentais, tanto na visão de alunos quanto de professores e com os resultados da Ref.~\cite{Alves} que afirma que o uso de atividades experimentais é de grande importância para aluno e contribui de forma significativa para os mesmos. 

%%%%%%%%%%%%%%%%%%%%%%%%%%%%%%%%%%%%%%%%%%%%%%%%%
%%%%%%%%%%%%%%%%%%%%%%%%%%%%%%%%%%%%%%%%%%%

\begin{figure}[!h]
\centering
\includegraphics[scale=0.6]{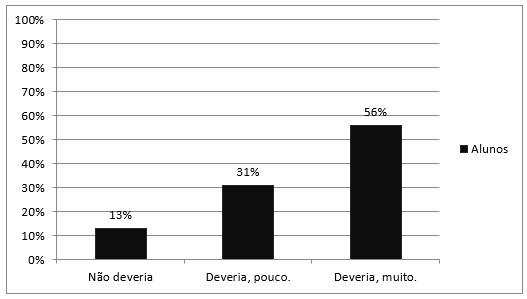}
\caption{Percentual de respostas quanto a frequência que o experimento deve ser usado nas aulas.}
\label{grafico2}
\end{figure}

Ao serem perguntados sobre a frequência com que os experimentos deveriam ser utilizados em aula, temos respostas bem satisfatórias e condizentes com já afirmado pelos alunos, indicando que as aulas com auxílio da atividade experimental chamam a atenção dos alunos. Estes resultados obtidos são mostrados no gráfico da Figura~\ref{grafico2} e se aproximam muito, 87\%, do resultado da Ref.~\cite{Dourado}, em que 90\% dos entrevistados responderam que gostariam que os professores de ciências (Física, Química e Biologia) utilizassem as TIC ou demostrando melhor os fenômenos estudados nestas áreas.

%%%%%%%%%%%%%%%%%%%%%%%%%%%%%%%%%%%%%%%%%%%%%%%%%
%%%%%%%%%%%%%%%%%%%%%%%%%%%%%%%%%%%%%%%%%%%

\begin{figure}[!h]
\centering
\includegraphics[scale=0.6]{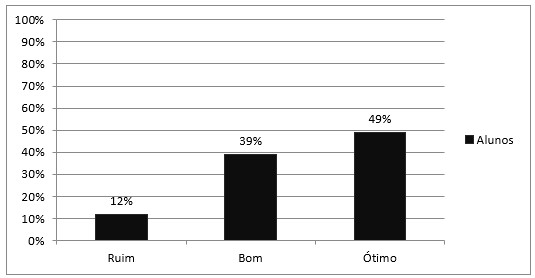}
\caption{Percentual de alunos que aprovam o Arduíno como uma ferramenta para produção de experimentos de física.}
\label{grafico3}
\end{figure}

Para avaliar a funcionalidade do Arduíno, a partir da discussão dos conceitos iniciais sobre o Arduíno, perguntou-se aos alunos sobre a dinamicidade da placa para produção de experimentos de física. A pergunta foi feita com o intuito de descobrir se o Arduíno é uma boa ferramenta metodológica para o ensino aprendizagem de física. Os dados são mostrados no gráfico da Figura~\ref{grafico3}.

Os dados do gráfico apresentado, demostram que o Arduíno é uma ótima ferramenta para utilização e produção do ensino de física, já que 88\% dos entrevistados aprovam o mesmo. Desse modo o Arduíno se torna uma ferramenta inovadora na prática educacional do ensino de física e as demais ciências.

%%%%%%%%%%%%%%%%%%%%%%%%%%%%%%%%%%%%%%%%%%%%%%%%%
%%%%%%%%%%%%%%%%%%%%%%%%%%%%%%%%%%%%%%%%%%%

Em relação aos experimentos apresentados em sala de aula (Termômetro e Absorção Luminosa), os alunos relatam como estes experimentos contribuíram para as aulas. Os mesmos destacam que o experimento contribuiu deixando as aulas mais dinâmica e interativa. Nas palavras do aluno \textbf{A}, “\textit{a aula fica mais diversificada e interativa}”, enquanto do aluno \textbf{B} relata que “\textit{com o experimento fica muito mais fácil o entendimento do conteúdo}”. Outros alunos destacam que “ \textit{os experimentos deixaram a aula mais dinâmica}”. Essas respostas nos mostram que o experimento diversifica a aula e deixa a mesma mais atraente aos olhos dos alunos.

Durante a aula, os alunos demonstraram muito interesse acerca do experimento. Durante a apresentação do experimento de termologia, os alunos questionaram o professor a respeito do uso de outros objetos com diferentes temperaturas para fazer aumentar ou diminuir a temperatura de outro corpo. O aluno \textbf{C}, questionou: “\textit{Professor, quando chegamos próximo do sensor há uma variação de temperatura, nesse caso nosso corpo está transferindo calor?} ”. Questionamentos como esse nos mostra que o experimento trouxe à tona o sentido observador do aluno, onde mesmo por meio da observação experimental concluiu que o corpo humano deve emitir calor. O aluno \textbf{D}, conversando com seus colegas de grupo, discute sobre a variação de temperatura em corpos com cores diferentes dizendo, “\textit{engraçado, quando o sensor é submetido à luz do celular ele varia menos graus em certo tempo do que quando submetido à proximidade da água em estado sólido}”. 

Acerca do experimento de óptica, os alunos discutiram sobre experiências do dia a dia e comparam com o observado no experimento, os mesmos comentam sobre o uso de camisas de cor escura e camisas de cor clara na presença do Sol e fazem o paralelo com o observado no experimento, nas palavras do aluno \textbf{E}: “\textit{Professor, o que acontece no experimento é o mesmo que quando usamos camisas de cor preta no Sol e esquenta mais do que quando usamos camisas de outra cor?}”, esse questionamento demostrando que os mesmos entenderam o conteúdo sobre absorção de radiação em diferentes cores e conseguiu levar o entendimento para a realidade dele. Outros alunos questionam sobre a distância que os objetos estão da fonte de calor, o aluno \textbf{F}, pergunta: “\textit{Professor, se aumentamos a distância entre a lâmpada e as semiesferas, terá o mesmo aumento de temperatura em relação ao tempo?} ”. Dessa forma, observamos que a atividade experimental, gera no aluno um sentimento investigativo, deixando a aula mais participativa por parte do aluno.

\section{Considerações finais}

Em vista da necessidade de se estudar novas metodologias como já apresentado nas seções anteriores deste artigo, tivemos o objetivo de produzir experimentos de física, com o auxílio da plataforma Arduíno, que podem ser utilizados em sala de aula e desse modo investigar se os experimentos irão proporcionar ao estudante uma aula mais dinâmica e assim propor método didático competente para o auxílio do ensino aprendizagem de física. 

Sendo assim, diante do que foi exposto no presente trabalho, observa-se que a utilização de atividades experimentais no ensino de física se faz necessário, tendo em vista o aumento percentual de alunos que fizeram alguma prática experimental em laboratório, mesmo que as escolas em que estudam não tivessem promovido essas práticas, além dos experimentos proporcionarem aos estudantes um aprendizado significativo, já que mais de 50\% dos entrevistados responderam de forma favorável o uso de experimentos em sala de aula. Portanto, esses experimentos devem ser utilizados com mais frequência e o Arduíno se faz uma ótima ferramenta para produção de experimentos no ensino de física.

%%%%%%%%%%%%%%%%%%%%%%%%%%%%%%%%%%%%%%%%%%%%%%%%%%%%%%%%%%%%%%%%%%%%%%%%%%%%%%%%
\section*{Apêndice A: roteiro de atividade para o experimento de termologia}

MATERIAIS NECESSÁRIOS:

\begin{itemize}

\item Arduíno UNO Rev. 3;

\item Sensor de temperatura, umidade e pressão BMP 180;
 
\item 02 potenciômetro;

\item Tela de LCD;

\item Jumpers;

\item Protoboard;

\item Fonte ou cabo USB; 

\item 03 objetos com diferentes temperaturas.

\end{itemize}

DEVERÁ SABER AO FINAL DA ATIVIDADE:

\begin{itemize}

\item Conceituar temperatura;

\item Conceituar calor;

\item Conhecer o processo de equilíbrio térmico;

\item Conhecer escalas termométricas;

\item Descrever o gráfico e a função matemática da temperatura e função do tempo.

\end{itemize}

DESCRIÇÃO DA ATIVIDADE:

\begin{enumerate}

\item Ao aproximar o sensor dos diferentes corpos, o que acontece?

\item Preencha os dados da tabela (cf. Figura~\ref{tabela1}) para cada corpo utilizado.

%%%%%%%%%%%%%%%%%%%%%%%%%%%%%%%%%%%%%%%%%%%%%%%%%
\begin{figure}[!h]
\centering
\includegraphics[scale=0.55]{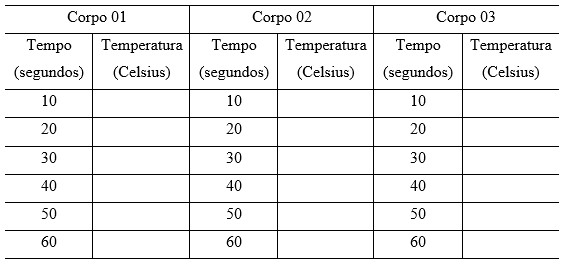}
\caption{Tabela referente aos experimento de termologia.}
\label{tabela1}
\end{figure}
%%%%%%%%%%%%%%%%%%%%%%%%%%%%%%%%%%%%%%%%%%%

\item Faça a análise dos dados contidos na tabela e responda os seguintes questionamentos para cada corpo:

\begin{enumerate}
\item[I] A temperatura variou ou permaneceu constante?

\item[II] A temperatura estava aumentando ou diminuindo?
\end{enumerate}

\item Por que ocorreu a variação de temperatura?

\item Para cada objeto (agua em estado sólido, lâmpada do celular e corpo humano), Construa o gráfico da variação de temperatura em função do tempo e indique a  qual tipo de função que descreve o gráfico.

\item Quais as principais escalas termométricas?

\end{enumerate}

%%%%%%%%%%%%%%%%%%%%%%%%%%%%%%%%%%%%%%%%%%%%%%%%%%%%%%%%%%%%%%%%%%%%%%%%%%%%%%%%%%

\section*{Apêndice B: roteiro de atividades para o experimento de óptica}

MATERIAIS NECESSÁRIOS:

\begin{itemize}

\item Arduino UNO;

\item 02 Sensores de temperatura, umidade e pressão BMP 180;

\item 02 potenciômetros;

\item Tela de LCD;

\item Jumpers;

\item Protoboard;

\item Fonte ou cabo USB;

\item Duas semiesferas metálicas com cores diferentes na parte interna;

\item Uma lâmpada incandescente.

\end{itemize}

DEVERÁ SABER AO FINAL DA ATIVIDADE:

\begin{itemize}

\item Reconhecer os processos de absorção da luz;

\item Entender que a luz é uma forma de energia que quando propagado para outro corpo se transforma em outra forma de energia;

\item Cores diferentes absorvem color de maneira diferente;

\item Conhecer os conceitos de Reflexão, absorção e transmissão da luz. 

\end{itemize}

DESCRIÇÃO DA ATIVIDADE:

\begin{enumerate}

\item Ao ligar a lâmpada, o que se pode observar nas placas?

\item Qual a principal diferença entre uma placa e a outra após a lâmpada ser ligada?

\item Preencha os dados da tabela (cf. Figura~\ref{tabela2}).

%%%%%%%%%%%%%%%%%%%%%%%%%%%%%%%%%%%%%%%%%%%%%%%%%
\begin{figure}[!h]
\centering
\includegraphics[scale=0.55]{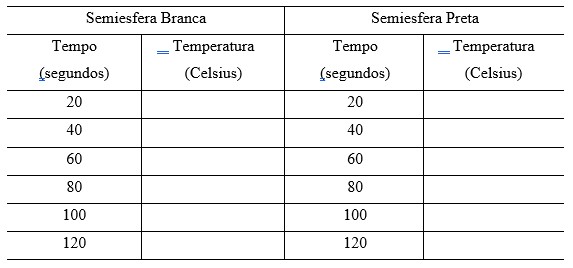}
\caption{Tabela referente aos experimento de óptica.}
\label{tabela2}
\end{figure}
%%%%%%%%%%%%%%%%%%%%%%%%%%%%%%%%%%%%%%%%%%%

\item A partir dos dados da tabela anterior, a que ser deve a variação ou não da temperatura nas placas?

\item Há diferença entre a temperatura das placas? Se sim, qual o motivo? 

\item Qual a relação entre a variação de temperatura e a cor das placas? 

\end{enumerate}

%%%%%%%%%%%%%%%%%%%%%%%%%%%%%%%%%%%%%%%%%%%%%%%%%%%%%%%%%%%%%%%%%%%%%%%%%%%%%%%%%%

\section*{Agradecimentos}

Os autores agradecem às escolas estaduais Landri Sales e a Escola Normal Oficial de Picos pela colaboração durante a realização da pesquisa; ao PIBIC-IFPI (Edital n. 57/2015) pelo suporte à pesquisa e a professora Maria Girlandia de Sousa pela leitura crítica e sugestões para o melhoramento da escrita do artigo.

%%%%%%%%%%%%%%%%%%%%%%%%%%%%%%%%%%%%%%%%%%%%%%%%%%%%%%%%%%%%%%%%%%%%%%%%%%%%%%%%%%

\end{document}